\documentclass[conference]{IEEEtran}
\usepackage{cite}
\usepackage{amsmath,amssymb,amsfonts}
\usepackage{algorithmic}
\usepackage{graphicx}
\usepackage{textcomp}
\usepackage{xcolor}
\usepackage{colortbl}
\usepackage{tabularx}
\usepackage{relsize}
\usepackage{pifont}
\usepackage{booktabs} 
\usepackage{multirow}
\usepackage{multicol}
\usepackage{adjustbox}
\usepackage{float}
\usepackage{changepage}
\usepackage{url}
\usepackage{enumitem}

\ifCLASSINFOpdf
\else
\fi
\begin{document}
%
\title{Quantitative Benchmarking of Anomaly Detection Methods in Digital Pathology}

\author{Can Cui$^{1}$, Xindong Zheng$^{1}$, Ruining Deng$^{1,2}$, Quan Liu$^{1}$, Tianyuan Yao$^{1}$, Keith T Wilson$^{3,4}$, Lori A Coburn$^{3,4}$, Bennett A Landman$^{1,5}$, Haichun Yang$^{6}$, Yaohong Wang$^{7}$, Yuankai Huo$^{1,5,6}$*}

\author{\IEEEauthorblockN{Can Cui}
\IEEEauthorblockA{Department of Computer Science\\
Vanderbilt University, TN, USA}
\and
\IEEEauthorblockN{Xindong Zheng}
\IEEEauthorblockA{Department of Computer Science\\
Vanderbilt University, TN, USA}
\and
\IEEEauthorblockN{Ruining Deng}
\IEEEauthorblockA{Department of Radiology\\ Weill Cornell Medicine, NY, USA}
\and
\IEEEauthorblockN{Quan Liu}
\IEEEauthorblockA{Department of Computer Science\\
Vanderbilt University, TN, USA}
\and
\IEEEauthorblockN{Tianyuan Yao}
\IEEEauthorblockA{Department of Computer Science\\
Vanderbilt University, TN, USA}
\and

\IEEEauthorblockN{Keith T Wilson}
\IEEEauthorblockA{Department of Medicine\\
Vanderbilt University Medical Center, TN, USA\\
Veterans Affairs Tennessee Valley\\ Healthcare System TN, USA}
\and
\IEEEauthorblockN{Lori A Coburn}
\IEEEauthorblockA{Department of Medicine\\
Vanderbilt University Medical Center\\
Veterans Affairs Tennessee Valley\\ Healthcare System TN, USA}
\and
\IEEEauthorblockN{Bennett
A Landman}
\IEEEauthorblockA{Department of Electrical and\\ Computer Engineering\\Department of Computer Science\\
Vanderbilt University, TN, USA}
\and
\IEEEauthorblockN{Haichun Yang}
\IEEEauthorblockA{Department of Pathology, Microbiology,\\ and Immunology\\ Vanderbilt University Medical Center, TN, USA}
\and
\IEEEauthorblockN{Yaohong Wang}
\IEEEauthorblockA{Department of Anatomical Pathology\\ UT MD Anderson Cancer Center,TX, USA}
\and
\IEEEauthorblockN{Yuankai Huo}
\IEEEauthorblockA{Department of Computer Science\\Department of Electrical and Computer Engineering\\
Vanderbilt University, TN, USA\\
$\star$ Corresponding Author}}


%


\maketitle

\begin{abstract}
Anomaly detection has been widely studied in the context of industrial defect inspection, with numerous methods developed to tackle a range of challenges. In digital pathology, anomaly detection holds significant potential for applications such as rare disease identification, artifact detection, and biomarker discovery. However, the unique characteristics of pathology images—such as their large size, multi-scale structures, stain variability, and repetitive patterns—introduce new challenges that current anomaly detection algorithms struggle to address. In this quantitative study, we benchmark over 20 classical and prevalent anomaly detection methods through extensive experiments. We curated five digital pathology datasets—both real and synthetic—to systematically evaluate these approaches. Our experiments investigate the influence of image scale, anomaly pattern types, and training epoch selection strategies on detection performance. The results provide a detailed comparison of each method's strengths and limitations, establishing a comprehensive benchmark to guide future research in anomaly detection for digital pathology images.
\end{abstract}


%
\IEEEpeerreviewmaketitle

\section{Introduction}

\begin{figure}
\begin{center}
\includegraphics[width=1\linewidth]{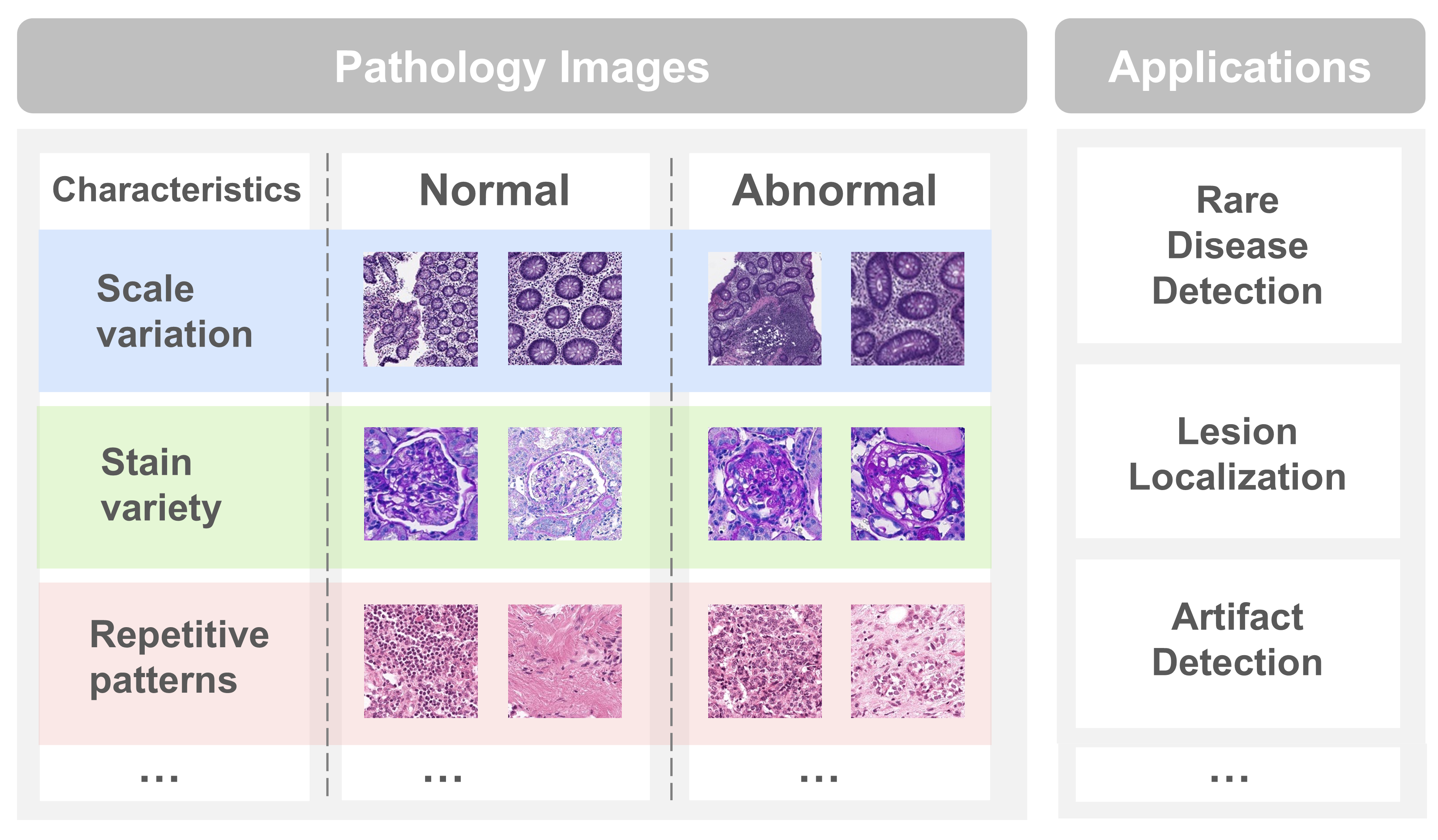}
\end{center}
   \caption{\textbf{Key characteristics and clinical applications of anomaly detection in digital pathological images.} Compared to natural images, anomaly detection in pathological images has the following characteristics: 1) Greater scale variation in multiscale, including cells and tissues. 2) Diversity in staining, such as differences in staining intensity and imaging brightness. 3) Repetitive patterns, with blurry boundaries between normal and abnormal.  
}
\label{fig: Overview}
\end{figure}

\sloppy
Anomaly detection refers to a model's ability to identify deviations from the learned distribution of training data during inference. These models are typically trained solely on normal data but distinguish between normal and abnormal data during inference, which is categorized as unsupervised/semi-supervised anomaly detection~\cite{liu2024deep, nassif2021machine}. This paradigm is particularly suitable for scenarios where normal data is abundant, while anomalies are rare, difficult to obtain during training, or highly variable with unknown manifestations. Anomaly detection has been widely applied in real-world tasks, including defect detection in manufacturing, abnormal event recognition in surveillance videos, and fraud detection in financial transactions.

Anomaly detection also has potential applications in the field of medical imaging. Pathology images are typically large-scale, with normal samples readily available from healthy individuals. Lesion areas are often small relative to normal regions, exhibit high pattern variability with unclear boundaries, and are rarely annotated by experts~\cite{gu2023using}. Additionally, privacy concerns may limit the availability of certain patient data, and some lesions may be extremely rare or entirely novel. By training only on normal data, anomaly detection can enable automatic lesion identification, with pixel-wise or patch-wise anomaly scores enhancing interpretability in disease diagnosis. Also, anomaly detection plays a crucial role in preclinical research, such as identifying the toxicological effects of candidate drugs~\cite{zingman2024learning, shelton2024automated}. Given the diverse manifestations of pathological abnormalities, supervised classification datasets often fail to capture all possible variations. In contrast, the availability of large-scale normal data allows anomaly detection methods to generalize to unseen anomalies in pathology images, enabling robust and broad applications.

However, pathology images differ significantly from radiology and natural images, posing unique challenges for anomaly detection Fig.~\ref{fig: Overview}. For example, pathology image analysis requires multi-scale interpretation, as pathologists frequently zoom in and out to capture both local and global diagnostic information. Staining variation also introduces additional complexity, requiring models to exhibit robustness to these differences~\cite{zingman2024learning}. Furthermore, pathology images often contain repetitive patterns, while anomalous regions tend to be highly complex and diverse, reducing the effectiveness of conventional anomaly detection algorithms. Therefore, we aim to evaluate the performance of general anomaly detection methods on pathology images.

This quantitative study focuses on benchmarking anomaly detection in pathology images. We summarize prior work, comprehensively benchmark 23 anomaly detection methods on real digital pathology images and synthetic datasets, and discuss the challenges and future directions in this field. The contributions of this work are summarized as follows:

\begin{itemize} \item We benchmark 23 classic and state-of-the-art anomaly detection methods on synthetic and real digital pathology images, featuring distinct anomaly patterns for comprehensive evaluation. \end{itemize}

\begin{itemize} \item Our experiments explore the impact of key components on anomaly detection performance in digital pathology, including image scale, reversed anomaly patterns, and strategies for selecting training epochs, which are rarely discussed in previous studies. \end{itemize}

\begin{itemize} \item We review the application of existing anomaly detection methods to pathology images and, combined with our experimental results, provide valuable insights to guide future research in this field. \end{itemize}

\section{Related work}

Several reviews summarize anomaly detection in industrial device images~\cite{liu2024deep, cui2023survey, cao2024survey}. In recent years, there have been some review papers focused on anomaly detection in medical imaging~\cite{cai2025medianomaly, lagogiannis2023unsupervised, bao2024bmad}. The most recent one, Cai et al \cite{cai2025medianomaly}., benchmarked several medical datasets with different modalities and conducted extensive experiments to analyze the characteristics of different methods, including the impact of pretrained weights, loss functions, and key hyperparameter settings. Similarly, Lagogiannis et al. \cite{lagogiannis2023unsupervised} and Bao et al \cite{bao2024bmad} also benchmarked anomaly detection methods on various medical images and experimentally examined the factors such as pretrained weights, the size of the dataset, model efficiency, etc.  In our work, we selected a set of widely adopted methods from recent years in both industrial and medical image anomaly detection. These include methods from two publicly available GitHub benchmark repositories (mentioned in the experiment section), as well as additional milestone or state-of-the-art approaches. Many of these methods also overlap with those covered in previous review papers. However, most of these reviews primarily focus on radiology images, overlooking the unique characteristics of pathology images, such as their scale, large image sizes, and repetitive patterns.

Compared to previous work, our contributions are as follows: 1) Reviewing anomaly detection methods specifically applied to pathology images. 2) Summarizing the domain-specific characteristics of pathology image anomaly detection, and introducing real and synthetic datasets tailored for analysis. 3) Exploring a practical yet challenging problem—unbiased training epoch selection—which has not been systematically examined in prior work. 4) Applying representative benchmark methods to pathology image anomaly classification tasks, providing empirical insights into their performance in this domain.

\section{Anomaly Detection Algorithm}

\subsection{Anomaly detection in vision domain}
The strict anomaly detection algorithm we discuss here can only access normal data during the training phase, which is denoted as \( D_{\text{train}} = \{X_{\text{normal}}\} \). The goal of an anomaly detection algorithm is to learn a function \( \alpha(x, \theta) \) that distinguishes \( X_{\text{normal}} \) from \( X_{\text{abnormal}} \). Typically, a projection head \( \Phi \) is also introduced to convert the result of \( \alpha(x, \theta) \) into a numeric result, known as the anomaly score. This anomaly score is used to classify \( X_{\text{normal}} \) and \( X_{\text{abnormal}} \) in the test set. Generally, it holds that \( \Phi(\alpha(x_{\text{normal}}, \theta)) < \Phi(\alpha(x_{\text{abnormal}}, \theta)) \).

Since the training set does not include abnormal data, the core idea of the algorithm is to learn the \textbf{feature distribution of normal data} or the \textbf{behavior of some tasks (e.g., reconstruction, denoising) on normal data}. The feature distribution of unseen abnormal data or its behavior in a specific task is expected to differ from that of normal data, thereby enabling classification during the inference phase.

\begin{figure}
\centerline{\includegraphics[width=\columnwidth]{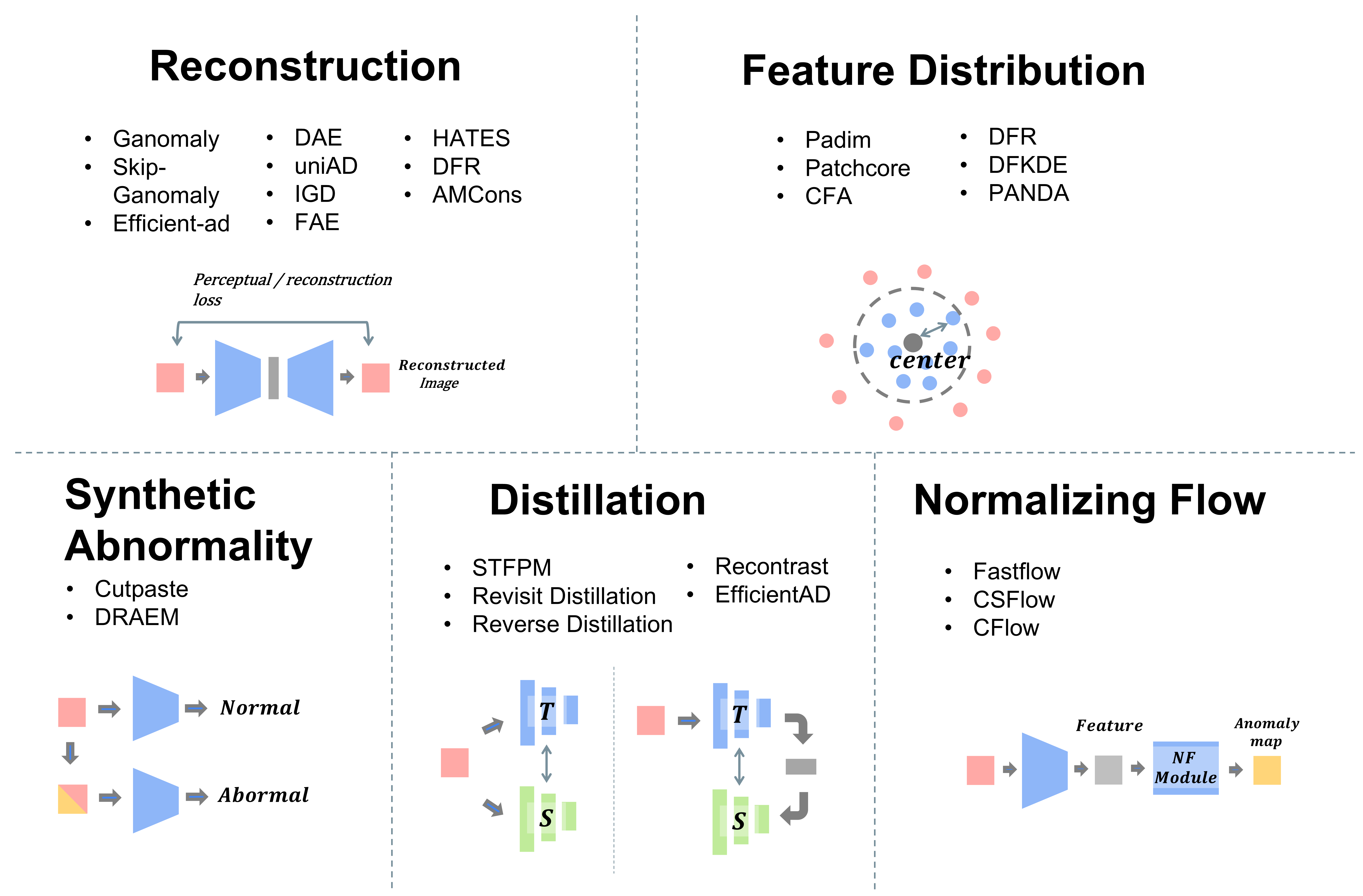}}
\caption{\textbf{Classification of anomaly detection models across five methodological families: reconstruction-based, feature distribution-based, distillation-based, normalizing flow-based, and synthetic anomaly-based methods.} Each family represents a distinct algorithmic principle for distinguishing anomalies from normal data. Reconstruction-based methods rely on detecting pixel-level differences, feature distribution methods model the latent space of normality, distillation methods leverage teacher–student discrepancies, normalizing flows model exact likelihood distributions, and synthetic approaches train models on augmented anomalies. These groupings are important as they highlight varying strengths and limitations when applied to the complex structure of pathology images.}

\label{fig: methods}
\end{figure}

\begin{table*}[h!]
\centering
\begin{adjustwidth}{0 pt}{0 pt}
\caption{Summary of anomaly detection methods categorized by type, with their main ideas, techniques, advantages, limitations, and representative works.}
\label{tab:anomaly_methods}
\begin{tabular}{|p{1.7cm}|p{3.6cm}|p{2.0cm}|p{3.0cm}|p{3.0cm}|p{2.4cm}|}
\hline
\textbf{Method Type} & \textbf{Main Idea} & \textbf{Techniques} & \textbf{Advantages} & \textbf{Limitations} & \textbf{Representative Work} \\ 
\hline
Image Reconstruction & Train models to reconstruct normal images and use reconstruction errors to identify anomalies. & AE/VAE, GAN, Diffusion models & Offers visual interpretability through reconstruction error maps. & May reconstruct abnormal data well; sensitive to model architecture (e.g., bottleneck size). & UniAD~\cite{you2022unified},~\cite{zavrtanik2021draem}, IGD~\cite{chen2022deep},~\cite{kascenas2022denoising}, AMCons~\cite{silva2022constrained}, GANomaly~\cite{akcay2019ganomaly}, FAE~\cite{meissen2022unsupervised}, DFR~\cite{yang2020dfr}, HATES~\cite{ghorbel2022transformer} \\ 
\hline
Feature Distribution & Model the distribution of feature embeddings extracted from normal data; anomalies deviate from this distribution. & Memory Bank, One-Class SVM, KNN, Clustering, GMM & Works well with high-dimensional data; flexible integration with pre-trained models. & Sensitive to noise and sparsity in embeddings; requires meaningful features. & PANDA~\cite{reiss2021panda}, PatchCore~\cite{roth2022towards}, CFA~\cite{lee2022cfa}, DFKDE~\cite{anomalib_dfkde},~\cite{ahuja2019probabilistic}, PaDiM~\cite{defard2021padim} \\ 
\hline
Distillation & Use knowledge distillation to train a compact model to mimic a teacher model; discrepancies are treated as anomalies. & Knowledge distillation, Feature alignment & Efficient inference with smaller model size. & Performance depends on teacher quality and distillation design. & Reverse Distillation~\cite{deng2022anomaly}, Revisit Distillation~\cite{tien2023revisiting}, STFPM~\cite{wang2021student}, ReContrast~\cite{guo2023recontrast} \\ 
\hline
Normalizing Flow & Learn exact data distributions using invertible transformations; use likelihood scores for anomaly detection. & Flow-based likelihood estimation & Expressive modeling of complex distributions. & Computationally expensive, especially on large-scale data. & FastFlow~\cite{yu2021fastflow}, CFLOW-AD~\cite{gudovskiy2022cflow}, CS-Flow~\cite{rudolph2022fully} \\ 
\hline
Synthetic Anomaly Detection & Introduce synthetic anomalies during training; train the model to distinguish them from normal data. & Self-supervised learning, Data augmentation & Adaptable to various anomaly types through creative augmentations. & Effectiveness depends on the quality/diversity of synthetic anomalies. & CutPaste~\cite{li2021cutpaste}, DRAEM~\cite{zavrtanik2021draem} \\ 
\hline
\end{tabular}
\end{adjustwidth}

\end{table*}

The reviewed anomaly detection algorithms are mentioned and categorized into different types, as shown in Fig.~\ref{fig: methods}, including reconstruction-based method, feature distribution method, distillation method, synthetic abnormality and normalizing flow method. The characteristics of these methods are summarized in Table~\ref{tab:anomaly_methods}. 

Reconstruction-based methods aim to reconstruct input images and flag discrepancies as anomalies based on reconstruction errors. Simpler architectures, such as autoencoders (AE/VAE), compress and reconstruct images. For example, IGD~\cite{chen2022deep} proposes fusing latent features within a Gaussian-distributed latent space to learn a smoother and more representative normal space. The HATES~\cite{ghorbel2022transformer} uses transformer blocks for autoencoder-based reconstruction, while FAE~\cite{meissen2022unsupervised} adopts the Structural Similarity Index Measure (SSIM) as an optimization objective. However, the bottleneck structure in autoencoders limits the expressiveness of abnormal patterns and simultaneously discards normal pattern information, leading to blurry reconstructions. Increasing the bottleneck size may help retain more information but risks also reconstructing anomalous content. To address this, DAE~\cite{kascenas2022denoising} introduces denoising strategies to recover meaningful signals and mitigate over-compression. Similarly, UniAD~\cite{you2022unified} employs a feature jittering strategy to encourage the model to recover the correct representation from noisy inputs. In addition, generative adversarial networks (GANs), such as GANomaly, and diffusion-based methods, such as AnoDDPM, have been introduced to produce higher-quality reconstructions with clearer details.

Feature distribution-based methods operate in the latent feature space by modeling the distribution of embeddings extracted from normal data. Techniques such as PatchCore, PANDA, PaDiM, and DFKDE rely on pretrained backbone networks to estimate feature statistics. These methods are typically more robust to complex patterns, as they do not depend on pixel-level reconstruction fidelity. PaDiM~\cite{defard2021padim} and DFKDE~\cite{anomalib_dfkde} extract features using pretrained models and model them with multivariate Gaussian distributions. PatchCore~\cite{roth2022towards} constructs a memory bank via core-set sampling and uses nearest neighbor voting to classify predictions as normal or abnormal. CFA~\cite{lee2022cfa} enhances the Patchcore by organizing image features on a hyperspherical feature space. PANDA learns a compact feature space representing normal patterns, with anomalies expected to fall outside this learned manifold.

Knowledge distillation-based methods adopt a teacher–student paradigm, where the student network learns to replicate the outputs or feature representations of a teacher network on normal data. A significant discrepancy between the student and teacher on abnormal samples indicates potential anomalies. Representative approaches include STFPM~\cite{wang2021student}, reverse distillation~\cite{deng2022anomaly}, ReContrast~\cite{guo2023recontrast}, and revisit distillation~\cite{tien2023revisiting}.

Normalizing flow-based methods model the likelihood of normal data through invertible transformations. These methods offer principled probabilistic anomaly scoring based on exact density estimation. However, they are often limited by high computational cost and sensitivity to architectural choices. CFLOW-AD~\cite{gudovskiy2022cflow} incorporates positional encoding alongside flow-based modules, while CS-Flow~\cite{rudolph2022fully} integrates multi-scale features for more accurate distribution estimation.

Finally, synthetic anomaly-based methods, such as CutPaste~\cite{li2021cutpaste} and DRAEM~\cite{zavrtanik2021draem}, simulate artificial anomalies through data augmentation or generative modeling. These methods enable training without real anomaly labels by teaching the model to distinguish between original and perturbed samples. For instance, CutPaste trains two separate encoders using a three-way classification task, distinguishing each input image from two variants with synthetic anomalies—designed for both image-level detection and pixel-level localization. The effectiveness of these methods heavily depends on the diversity and realism of the synthetic patterns, which may not always reflect true clinical abnormalities.

The methods outlined above represent the major families of anomaly detection techniques in the computer vision literature. In the following section, we review how anomaly detection methods have been adapted and applied to pathology images. In our study, we selected 23 representative algorithms spanning these five categories to systematically benchmark their performance in the domain of digital pathology.

\subsection{Anomaly detection in pathology images}

With the development of anomaly detection algorithms in the field of natural images, some studies have extended these algorithms to pathology images. Most of these approaches were based on reconstruction-based techniques. For instance, recent studies~\cite{shvetsova2021anomaly, zehnder2022multiscale, pocevivciute2021unsupervised} have utilized autoencoders to reconstruct normal images under the assumption that anomalies, unseen during training, will produce higher reconstruction errors. However, pixel-wise reconstruction scores can be overly sensitive to local artifacts such as edges or background regions, and may fail to capture meaningful global semantic discrepancies. To address this, perceptual loss~\cite{johnson2016perceptual} is often incorporated as an auxiliary loss and anomaly scoring metric in these autoencoder-based methods.Despite this improvement, autoencoder-reconstructed images often remain blurry. To generate higher-quality reconstructions, Generative Adversarial Network (GAN)- based and diffusion-based methods have been introduced. GANomaly~\cite{akcay2018ganomaly} pioneered the application of GANs to anomaly detection, and~\cite{lai2024unsupervised} adapted this framework to pathology images. Similarly,~\cite{pocevivciute2021unsupervised} and~\cite{shelton2024automated} utilized StyleGAN as the generative backbone, while~\cite{gu2023using} leveraged Progressive GAN for producing high-resolution pathology images. More recently,~\cite{linmans2024diffusion} employed the diffusion-based AnoDDPM model for pathology image reconstruction. 

Feature distribution-based anomaly detection typically leverages the strong representational power of pre-trained models to distinguish between normal and abnormal images. Although its application in medical imaging, especially in the domain of pathology images, is less common than reconstruction-based approaches, it has demonstrated promising performance in several prior studies. For example,~\cite{zingman2024learning} employed a pre-trained model and an auxiliary classification task to train a classification backbone for extracting image features. They used central loss to learn a compact representation of normal features, which helps distinguish the dispersed distribution of abnormal features.

It is especially worth noting that many anomaly detection algorithms have undergone adaptations to better suit the complex nature of pathology images. For example, Gu et al.~\cite{gu2023using} and Shvetsova et al.~\cite{shvetsova2021anomaly} adopted progressive training strategies to capture the fine details of pathology images.~\cite{zehnder2022multiscale} introduced skip connections similar to U-Net architectures to retain more details. To further align feature extraction with pathology images, Zingman et al.~\cite{zingman2024learning} defined an auxiliary classification task of normal mouse liver tissue, which is the same image domain for anomaly detection. This approach makes pre-trained models on ImageNet more suitable for extracting features from mouse liver images. Considering stain variations of pathology images, this study also proposed using mix-up color augmentation to reduce color sensitivity while preserving sensitivity to structural differences. Additionally, Lai et al.~\cite{lai2024unsupervised} simulated anomalies such as enlarged cell nuclei or irregular tissue structures when adapting CutPaste and local magnification techniques during training. The scale of pathology images is another important factor. Both Lat et al.~\cite{lai2024unsupervised} and Shvetsova et al.~\cite{shvetsova2021anomaly} considered the impact of scale and extracted multi-scale features to improve anomaly detection performance.

As for anomaly score calculation, in addition to common metrics such as Mean Squared Error (MSE), other similarity measures like learned perceptual image patch similarity (LPIPS) and structural similarity index measure (SSIM), as well as perceptual loss, are used to evaluate the similarity between generated and original images. Different metrics may show consistency or complement each other. Many studies combine multiple metrics for anomaly score computation, but the choice and weighting of these metrics need to be adjusted based on specific applications.

Since whole-slide images (WSIs) in pathology are too large to process directly, they are typically divided into patches. Anomaly scores are computed for each patch, and the WSI-level result is obtained by averaging or taking the maximum score. The choice of aggregation strategy can also affect the results.~\cite{zingman2024learning} discussed other aggregation strategies that can be considered.

However, these studies only used a fixed number of training epochs and did not further discuss how to fine-tune this important parameter, especially when the validation set may also lack abnormal images. Our subsequent experiments focus on this often-overlooked issue. In addition, we conducted a more detailed analysis of the impact of multiscale features on anomaly detection in pathology images using both real and synthetic datasets.

\section{Data and Experiments}

\subsection{Data}

\begin{figure}
\centerline{\includegraphics[width=\columnwidth]{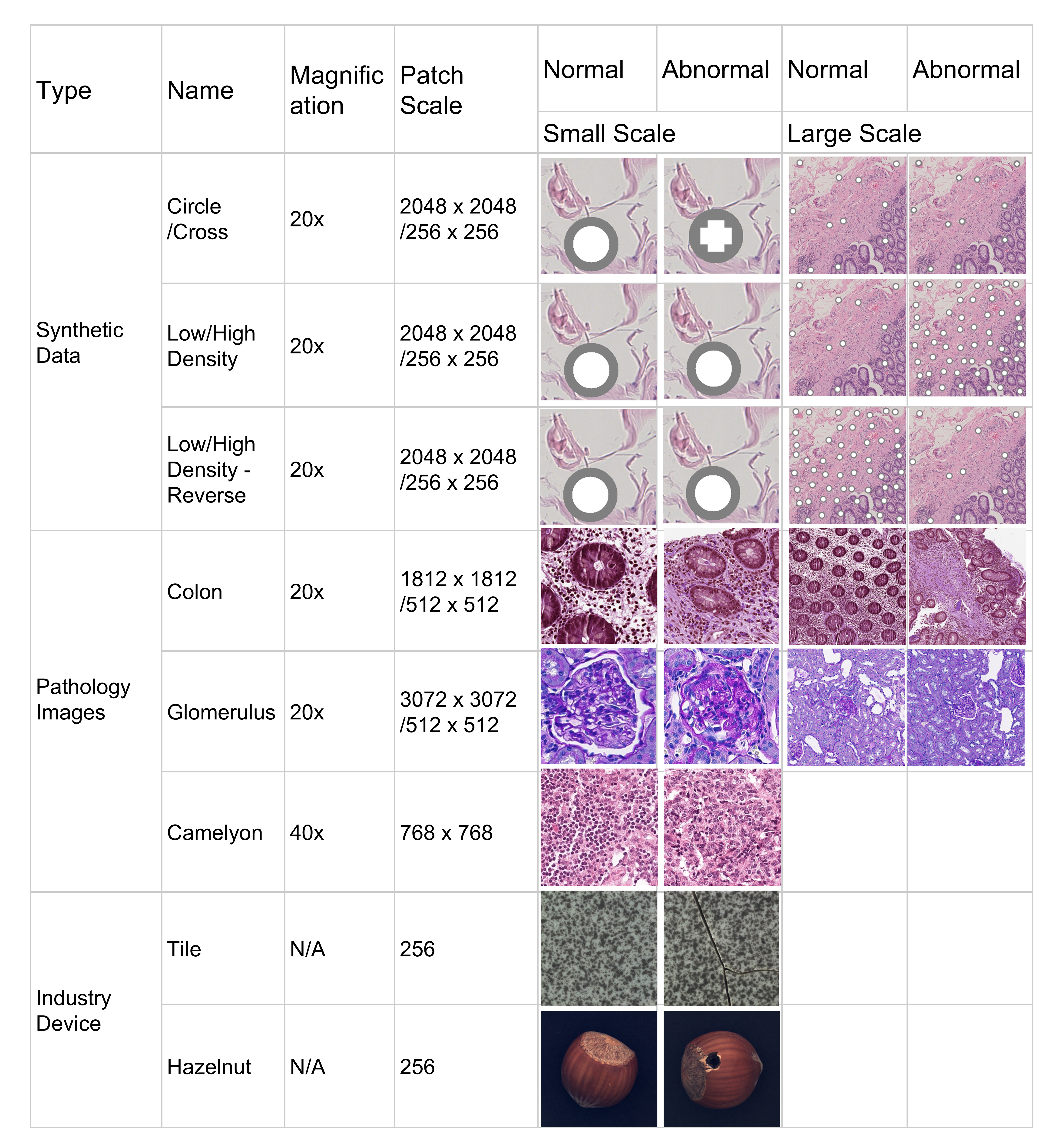}}
\caption{\textbf{Overview of the datasets and example images spanning diverse digital pathology and industrial imaging modalities.} The datasets include real-world pathology images (e.g., colon, glomerulus, Camelyon), synthetic pathology images with designed patterns, and standard industrial datasets (e.g., Hazelnut, Tile). This breadth ensures comprehensive evaluation across different anomaly types, image scales, and domain complexities, providing robust insights into the generalizability of each detection method.}

\label{fig:datasets}
\end{figure}

The dataset characteristics are summarized in Table~\ref{tab:Datasets}. This study uses three real-world pathology image datasets, two synthetic datasets, and two industrial defect datasets to evaluate different anomaly detection algorithms. Each dataset is divided into training, validation, and testing sets. The training set contains only normal data, while the testing set includes both normal and abnormal data, sharing a similar distribution to the validation set. Examples of normal and abnormal images are provided in Fig.~\ref{fig:datasets}.

\begin{table}[t]
\centering
\caption{Datasets and the corresponding data splits used in this work}
\resizebox{\columnwidth}{!}{
\begin{tabular}{|l|l|r|rr|rr|}
\hline
\multirow{2}{*}{Dataset} &
  \multirow{2}{*}{Scale} &
  \multicolumn{1}{l|}{Training} &
  \multicolumn{2}{l|}{Validation} &
  \multicolumn{2}{l|}{Testing} \\ \cline{3-7} 
 &
   &
  \multicolumn{1}{l|}{Normal} &
  \multicolumn{1}{l|}{Normal} &
  \multicolumn{1}{l|}{Abnormal} &
  \multicolumn{1}{l|}{Normal} &
  \multicolumn{1}{l|}{Abnormal} \\ \hline
\multirow{2}{*}{Synthetic Dataset} &
  small scale &
  1536 &
  \multicolumn{1}{r|}{1024} &
  1024 &
  \multicolumn{1}{r|}{1024} &
  1024 \\ \cline{2-7} 
                            & large scale & 24   & \multicolumn{1}{r|}{16}   & 16   & \multicolumn{1}{r|}{16}   & 16   \\ \hline
\multirow{2}{*}{Colon}      & small scale & 3408 & \multicolumn{1}{r|}{1136} & 1264 & \multicolumn{1}{r|}{1280} & 1280 \\ \cline{2-7} 
                            & large scale & 213  & \multicolumn{1}{r|}{71}   & 79   & \multicolumn{1}{r|}{80}   & 80   \\ \hline
\multirow{2}{*}{Glomerulus} & small scale & 649  & \multicolumn{1}{r|}{71}   & 71   & \multicolumn{1}{r|}{69}   & 65   \\ \cline{2-7} 
                            & large scale & 649  & \multicolumn{1}{r|}{71}   & 71   & \multicolumn{1}{r|}{69}   & 65   \\ \hline
Camelyon                    & combined           & 5462 & \multicolumn{1}{r|}{2150} & 2169 & \multicolumn{1}{r|}{4000} & 817  \\ \hline
Hazelnut                    & combined           & 391  & \multicolumn{1}{r|}{9}    & 12   & \multicolumn{1}{r|}{31}   & 58   \\ \hline
Tile                        & combined           & 201  & \multicolumn{1}{r|}{29}   & 19   & \multicolumn{1}{r|}{33}   & 65   \\ \hline
\end{tabular}
}
\label{tab:Datasets}
\end{table}

\subsubsection{Pathology Images}
\textbf{Breast Pathology Dataset~\cite{litjens20181399}}. Detecting lymph node metastases is a key factor in breast cancer diagnosis. The Camelyon16 challenge~\cite{bejnordi2017diagnostic} offers a public dataset for metastasis detection in digital pathology. Its training set includes 110 tumor-positive and 160 normal H\&E stained whole slide images (WSIs), while the test set contains 50 tumor-positive and 80 normal WSIs, with pixel-level tumor annotations. Following prior work~\cite{shvetsova2021anomaly}, we preprocess the 40\texttimes magnification WSI data into 768\texttimes768 patches for training, validation, and testing. Tumor patches are extracted from annotated tumor regions, while normal patches come from non-tumor regions. We used the same test set as previous work, consisting of 4000 normal and 817 abnormal patches. For validation, we randomly selected 2169 tumor and 2150 normal patches from the WSIs in the Camelyon16 training set, leaving 5462 normal patches for training. In contrast, previous work used all 7612 normal patches across our training and validation for model training.

\textbf{Colon Pathology Dataset~\cite{bao2021cross}.} Biopsies of 16 Crohn’s disease (CD) patients and 53 healthy controls collected from the colon. The tissues were stained with H\&E and scanned at 20\texttimes magnification. Large patches in size 1812 $\times$ 1812 were cropped from the WSIs and labeled by pathologists as either normal or abnormal (diseased). To explore the impact of different scales on anomaly detection, we further tiled the large patches into smaller patches of 512 $\times$ 512. Predictions on the small patches are aggregated and compared with predictions from the large patches to evaluate the performance across scale.

\textbf{Kidney Pathology Dataset~\cite{yao2022glo}}. This dataset consists of both benign and malignant glomeruli of mice. We extracted center-cropped patches with benign or malignant glomeruli at sizes of 3072 $\times$ 3072 and 512 $\times$ 512 separately from the WSIs stained by H\&E at 40$\times$ magnification. Normal patches are retained only if all glomeruli within them are normal.

\subsubsection{Synthetic Images}
Investigating abnormal patterns in pathology images presents significant challenges due to their complexity. So, we incorporated synthetic datasets to analyze model performance in a controlled environment. By utilizing 56 patches in 2048 $\times$ 2048 pixels tiled from 20$\times$ normal WSIs in the Unitopatho dataset~\cite{barbano2021unitopatho}, we overlay synthetic patterns designed specifically for anomaly detection. This approach enables the simulation of various scenarios pertinent to digital pathology.

\textbf{Circle or Cross Patterns.} To synthesize the morphological difference of normal and abnormal tissue, we introduced gray circles with a radius of 60 pixels on 2048 $\times$ 2048 patches and designed two distinct patterns: one consisting of white circles positioned within the gray circles, and the other featuring white crosses instead of circles. 

To assess the impact of different magnification levels, we prepared two scales of abnormality. At the 20$\times$ magnification (large scale), we resized the 2048 $\times$ 2048 patches to 256 $\times$ 256 for analysis. For the small-scale patches, we tiled the above 2048 $\times$ 2048 patches into small patches of size 256 $\times$ 256. The prediction results from the small scale will be aggregated and compared with those from the large scale. Distinguishing between the circle and cross patterns is relatively straightforward on the small scale, while it presents greater difficulty on the large scale.


\textbf{Density Patterns.} Cell density is another critical component in anomaly detection in digital pathology images. We continued to use the concentric circles mentioned previously as the primary element but modified their density within the 2048$\times$2048 patches. In this setup, we divided the patches into 256$\times$256 grids and assigned probabilities for each grid containing a circle, with one pattern reflecting a 75\% probability and another reflecting a 25\% probability.

Similar to the previous experiment, we prepared two scales of patches for this dataset. Theoretically, the large-scale observations should facilitate easier detection of density differences compared to those observed at the small scale.


\subsubsection{Industrial Device Images}

We also incorporated two classes of industrial defect anomaly detection datasets~\cite{bergmann2019mvtec}, which are commonly used benchmarks in the natural image domain. Specifically, we utilized the \textbf{Hazelnut dataset} to represent object abnormalities and the \textbf{Tile dataset} to capture texture abnormalities. These datasets serve as benchmarks, allowing us to illustrate the gap between the anomaly detection of industrial device images and medical images. As a validation set is not available in this dataset, we extracted a small portion of the testing data as the validation set in our experiment. 

\subsection{Experiment Settings}
In our experiments, the validation set is prepared in two configurations: a complete version with both normal and abnormal data, and a strict, unbiased version with only normal data. We analyze and compare different model selection strategy under these two validation setups in our experiments. Additionally, as detailed further in this section, we explore the impact of different pathology image scales and the reversion of abnormal patterns on model performance. Anomaly detection algorithms are applied to the prepared datasets described above, with three sets of experiments conducted for analysis.

\subsubsection{Impact of Different Scales}
For the in-house colon dataset, in-house glomerulus dataset, and the synthetic datasets, we do anomaly detection in different scales. The patches in a smaller scale were tiled from the same region in the original WSIs and contained more detail than the larger scale. The integration of the small region is used as the final prediction results to compare with the prediction of the larger scale. 

\subsubsection{Unbiased Training Epoch Selection}
An important yet often overlooked challenge in anomaly detection is selecting the appropriate training epoch for the model. In most cases, only normal data is available during the training phase, leaving no complete validation set to determine the optimal epoch as in supervised training. Even when limited abnormal data are available, they may lack the diversity and generality needed to represent all possible abnormal scenarios, potentially introducing bias. Furthermore, stopping the training process at different epochs can lead to significantly different outcomes, yet previous research has not sufficiently addressed the criteria for selecting training epochs. Following the work of~\cite{cui2023feasibility}, we implement four training epoch selection methods, including both with and without the inclusion of abnormal data during the training phase, to facilitate a broader comparison. Specifically, these four methods are:
\begin{itemize} 
    \item \textbf{Strategy 1 - Normal Sample Loss Evaluation:} This method, proposed by~\cite{reiss2021panda}, involves saving multiple model checkpoints from different epochs during training. During the testing phase, the anomaly score for each sample is calculated at various checkpoints and normalized by the corresponding average anomaly score of normal samples in the validation set, resulting in a value known as the maximal ratio. The checkpoint with the highest ratio indicates the best separation for that testing sample, and this maximal ratio serves as the anomaly score for the sample. The rationale behind this approach is that the maximum deviation ratio of a sample during the testing phase, compared to normal data in the validation set, reflects how likely the sample is to be abnormal. With this method, the number of epochs used to calculate the anomaly score may vary across different samples.
    \item \textbf{Strategy 2 - Complete Validation Set Method:} This method utilizes a complete validation set containing both normal and abnormal data. The epoch with the highest Area Under the Curve (ROC-AUC) score on this validation set is selected. If the data distribution of the validation set closely resembles that of the testing set, this method can serve as the upper bound for model selection performance as supervised learning.
    \item \textbf{Strategy 3 - Normal Sample Loss Evaluation:} This strategy evaluates the loss of normal samples in the validation set to determine how effectively the model has been trained on the pretext task, such as image reconstruction. The point of minimal loss indicates the best-trained model, which is then used for testing.
    \item \textbf{Strategy 4 - Last Epoch Selection:} This strategy assumes that training has either converged or the model has overfitted to the training samples. In this case, the model from the final epoch is directly selected for testing. As highlighted in~\cite{reiss2021panda}, continual learning can help ensure stable training, making the last epoch a reasonable choice.
\end{itemize}

\subsubsection{Reversion of the normality and abnormality}
The difficulty of learning diverse normal patterns can vary significantly. To investigate this, we reversed the definitions of normal and abnormal in the synthetic density dataset by swapping the normal and anomaly patterns. We then repeated the experiments to analyze the impact of this reversal.

The implementations of CS-Flow~\cite{rudolph2022fully}, CFA~\cite{lee2022cfa}, DFM~\cite{ahuja2019probabilistic}, DFKDE~\cite{anomalib_dfkde}, DRAEM~\cite{zavrtanik2021draem}, EfficientAD~\cite{anomalib_dfkde}, FastFlow~\cite{yu2021fastflow}, GANomaly~\cite{akcay2018ganomaly}, PADIM~\cite{defard2021padim}, PatchCore~\cite{roth2022towards}, and STFPM~\cite{wang2021student} were sourced from the GitHub repository: \url{https://github.com/openvinotoolkit/anomalib}. Similarly, CFLOW-AD, DAE~\cite{kascenas2022denoising}, DFR~\cite{yang2020dfr}, FAE~\cite{meissen2022unsupervised}, HATES~\cite{ghorbel2022transformer}, and Reverse Distillation were obtained from another repository: \url{https://github.com/iolag/UPD\_study}. The remaining algorithms—SkipGanomaly~\cite{akccay2019skip}, Revisit Distillation, PANDA~\cite{reiss2021panda}, IGD~\cite{chen2022deep}, ReContrast~\cite{guo2023recontrast}, and UniAD~\cite{you2022unified}—were implemented using code from their respective official GitHub repositories.

For a fair comparison, we standardized the experimental setup as follows: input image sizes were unified to 256\texttimes 256, with 200 training epochs and a batch size of 16. Pretrained weights were loaded when available, and images were standardized using the default mean and standard deviation values for ImageNet. Data augmentation included horizontal flipping, while most of the original parameters were retained.

For anomaly detection algorithms, a single anomaly score was obtained for each patch. When aggregation was required for object-wise or patient-wise results, the average of the top 20\% highest anomaly scores was used as the final anomaly score. A higher anomaly score indicates a greater likelihood of abnormality. The ROC-AUC metric was employed to evaluate the performance of the anomaly detection algorithms. 
Here is a refined version:

A higher ROC-AUC indicates better separation and effectiveness of the method in detecting anomalies.

All experiments were conducted using an NVIDIA RTX 4090 GPU.

\section{Results and Discussion}
\subsection{Results Analysis}
\begin{table*}
\centering
\begin{adjustwidth}{0 pt}{0 pt}
\caption{Experiment Results (ROC-AUC): This table presents the performance of various anomaly detection methods under different epoch selection strategies across multiple datasets. The four strategies include: sample-wise best epoch selection (sample), best epoch based on the full validation set (val), epoch with the minimum training loss (loss), and the final epoch (last). The result highlighted in red indicates the highest ROC-AUC among all strategies, while blue denotes the second-highest. 
Key observations:
1) Overall, feature distribution–based anomaly detection methods outperform both distillation-based and reconstruction-based approaches.
2) Based on the average results for sample-wise selection across 20 datasets, last epoch selection consistently performs the worst (19 out of 20), and the best and second-best performances are achieved by val and sample-wise respectively in 15 out of 20 datasets. Overall, the ranking is: \text{val} $>$ \text{sample-wise} $>$ \text{loss} $>$ \text{last}.
3) Image scale significantly affects detection performance. Despite many methods incorporating multi-scale information, directly using image patches of different scales still leads to notable differences in results, especially important in pathology images where scale variation is large.
4) On synthetic density datasets, reversing the definitions of normal and abnormal yields different results. This suggests that the difficulty of modeling normal patterns also impacts anomaly detection performance.}
\label{tab:results}
\includegraphics[width=\linewidth]{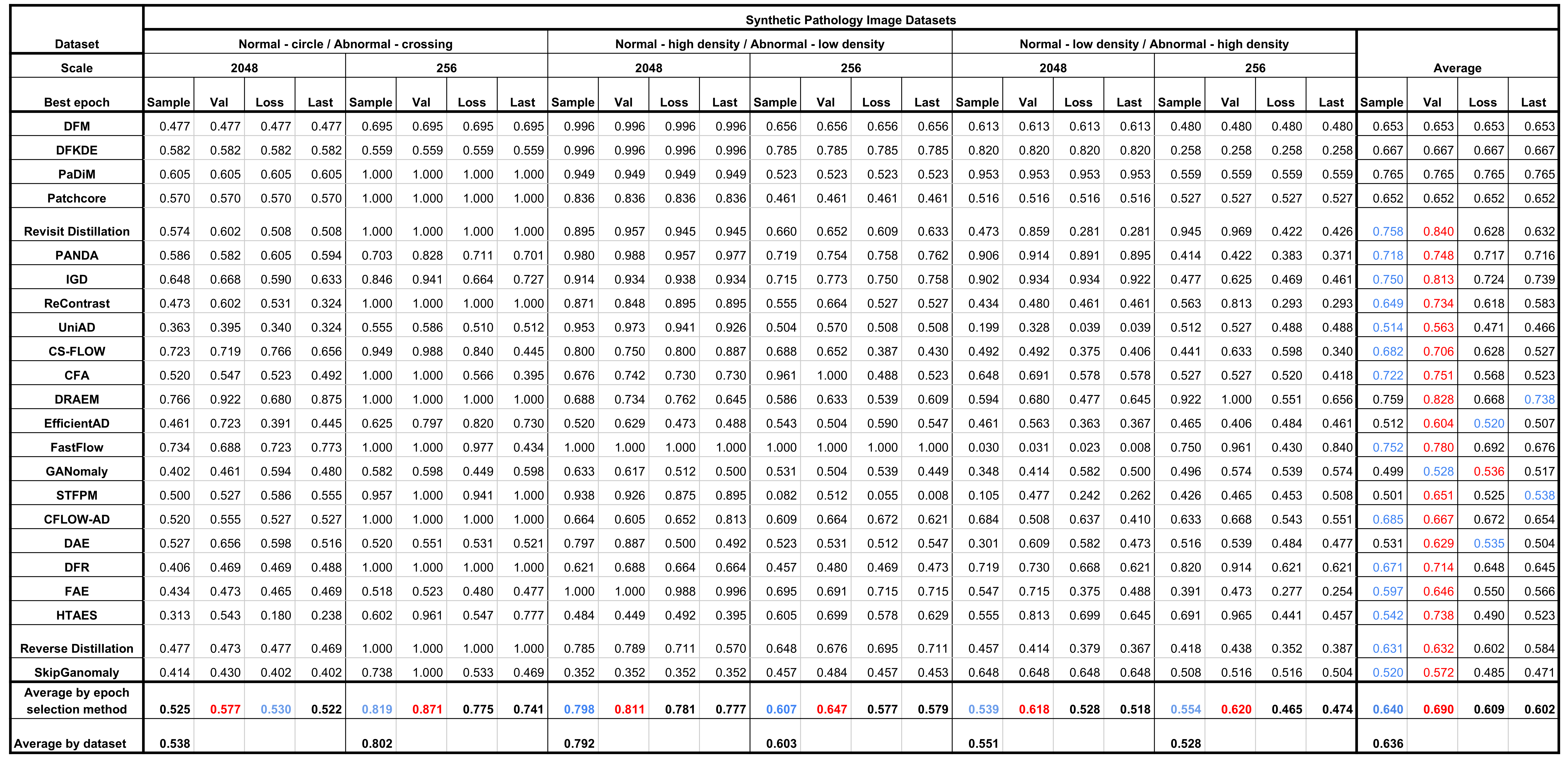}

\includegraphics[width=\linewidth]{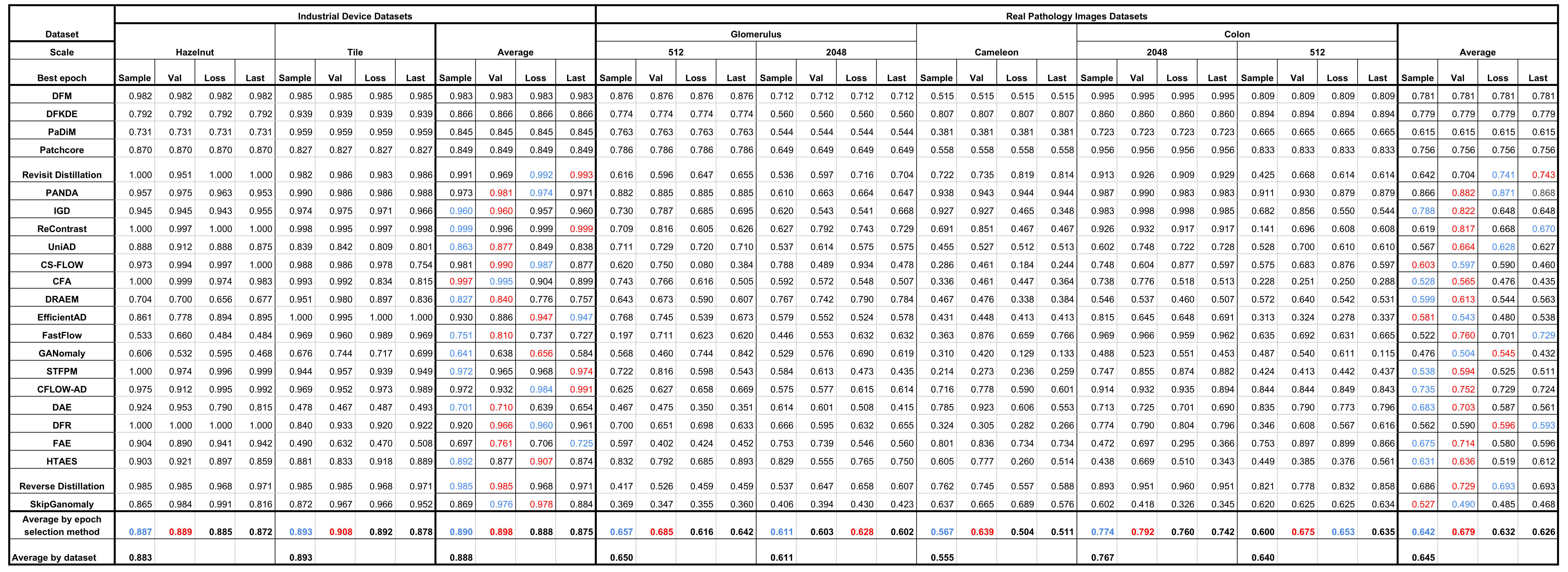}

\end{adjustwidth}

\end{table*}

\begin{table*}
\centering
\begin{adjustwidth}{0 pt}{0 pt}
\caption{Experiment Results (PR-AUC): it shows similar trends to ROC-AUC. Epochs selected using the full validation set achieved the best PR-AUC, while the loss-based selection performed similarly to sample-wise.}
\label{tab:results1}

\includegraphics[width=\linewidth]{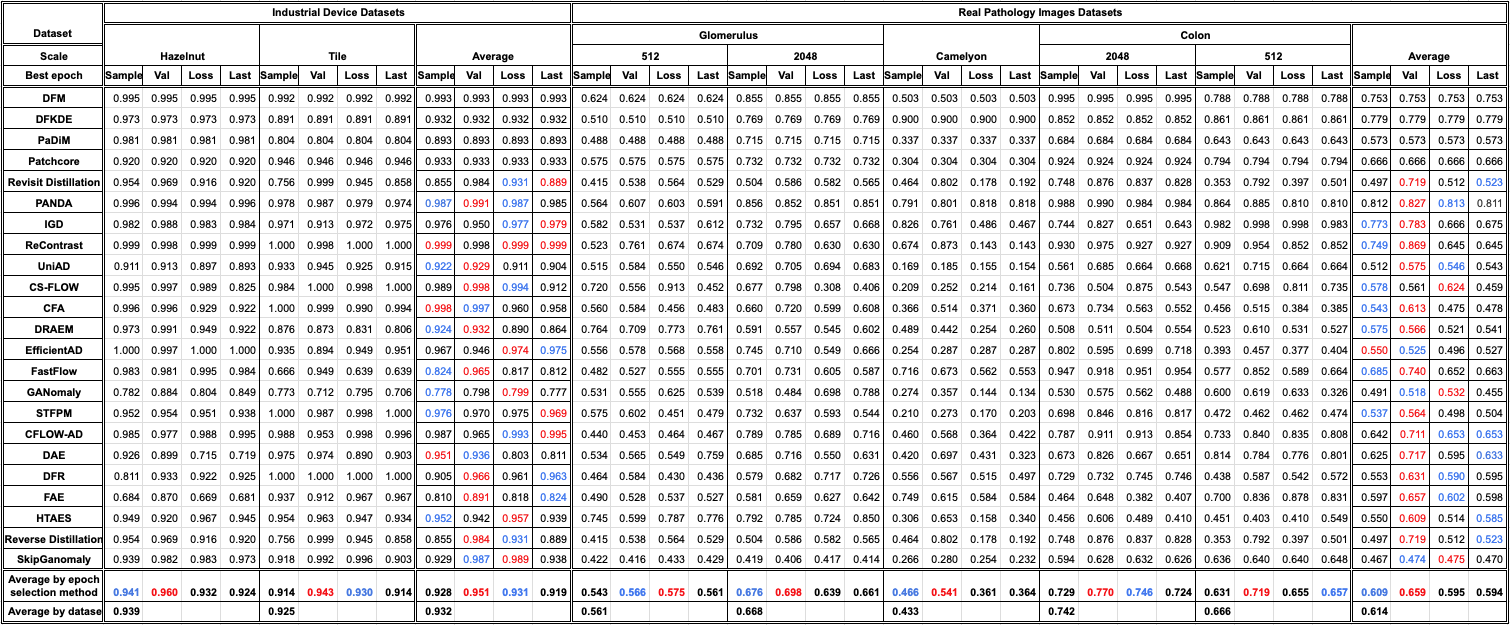}

\end{adjustwidth}

\end{table*}

Results of experiments are shown in Table~\ref{tab:results} and ~\ref{tab:results1}. 

\subsubsection{General Comparison among Different Methods and Datasets} There is no method that performs consistently the best among all datasets. This discovery also aligns with the conclusion in prior work~\cite{cui2023feasibility}. Meanwhile, some methods achieve competitive performance in some datasets (obviously above average) but almost fail (obviously below the average) in other datasets, for example Reverse distillation, skipGanomaly and uniAD, etc. Overall, feature distribution–based anomaly detection methods outperform both distillation-based and reconstruction-based approaches, achieving higher object-level ROC-AUC averages when using the complete validation set for epoch selection (distribution–based approaches: 0.692, distillation-based: 0.600, reconstruction-based approaches: 0.564). This trend is especially notable for pathology-type images and has also been observed in \cite{lagogiannis2023unsupervised}. Reconstruction-based methods are often limited by the unpredictable generalization. Sometimes anomalous images are reconstructed well, while some normal images are rendered overly blurry. Given the abundance of fine-grained patterns in pathology images, feature-based anomaly detection is often more robust than pixel-based error or variance metrics. Unlike distillation-based methods, which compress features through a teacher–student framework, distribution-based approaches model the feature space extracted from normal images directly. These methods typically preserve richer detail in the original feature space, making them more sensitive to subtle or unstructured anomaly patterns. Also, the average ROC-AUC for industrial device datasets is significantly higher than that for synthetic datasets and much higher than for real pathology datasets. This highlights the complexity and challenges associated with pathology images, particularly real pathology images. Methods that perform well on industrial datasets may not be sufficient to effectively distinguish between normal and abnormal cases in digital pathology images.

\subsubsection{Impact of Different Scales} For synthetic data, as expected, smaller scales perform better for anomaly patterns like crossing and circles because the details are more distinct in smaller patches. In contrast, for density-related abnormalities, larger scales, which capture global information, outperform smaller scales that lack sufficient context to reflect density changes.
In the case of the Glomerulus dataset, which has patterns similar to synthetic data with crossing and circles, smaller scales also perform better (avg ROC-AUC of small scale: 0.650, avg scale of large scale: 0.611). However, for the colon dataset, larger scales yield superior performance (avg ROC-AUC of large scale: 0.767, avg scale of small scale: 0.640). These findings underscore the critical role of scale in anomaly detection. While many methods, such as distillation, PANDA, and PADIM, leverage features from different network layers to capture multi-scale information, these approaches are insufficient when compared to the significant scale differences present in pathology images.

\subsubsection{Unbiased Training Epoch Selection} Regarding training epoch selection strategies, the method using the complete validation set consistently achieves the best performance with the highest average ROC-AUC for all three kinds of datasets. The sample-wise strategy ranks second, followed by the loss-based method in third place. The last-epoch method, while the simplest, performs the worst on average. 20 datasets, last epoch selection consistently performs the worst (19 out of 20), and the best and second-best performances are achieved by val and sample-wise respectively in 15 out of 20 datasets.
Both the sample-wise and loss-based strategies are unbiased since they do not require abnormal samples. The sample-wise strategy performs slightly better in our experiments, but it demands more memory to save models at various checkpoints and requires additional computation time. 

We also report PR-AUC values in Table ~\ref{tab:results1}, which show a similar overall trend to ROC-AUC. Training epochs selected based on the full validation set generally achieved the highest PR-AUC, while the loss-based selection method also performed well and yielded results comparable to the sample-wise strategy.

PR-AUC ignores true negatives and focuses more on the precision and recall of the positive class. It is more suitable for imbalanced datasets where the positive class is rare and false positives are costly. However, except for the Camelyon, Hazelnet, and Tile datasets, the test sets in our study are relatively balanced in terms of positive and negative samples. Therefore, ROC-AUC remains the most informative and reliable evaluation metric in our context.

\subsubsection{Reversion of the Normality and Abnormality} For pattern reversion, it is observed that when high density is considered normal and low density is abnormal, the performance achieves an average of 0.792, significantly higher than the 0.551 observed when low density is considered normal. High-density patterns might be relatively easier to learn. Almost every small-scale patch contains a dot, and its absence strongly suggests an anomaly. Conversely, when low density is defined as abnormal, the model struggles more to distinguish it. The experiment results that the difficulty of learning different normal patterns can vary greatly, even when the testing data remains the same but with different definitions of normality and abnormality. Especially for reconstruction-based methods, some normal patterns may be difficult to learn, while some abnormal patterns may still be well reconstructed.

\subsection{Future Directions}

\textcolor{red}{\subsubsection{Suggestions for Clinical Usage}}
{Given the stringent safety and accuracy requirements in clinical settings, anomaly detection algorithms are more likely to serve as \textbf{decision support tools} rather than standalone diagnostic systems. Since anomalies in medical images are often undefined in advance, performance inconsistency across datasets is expected. In pathology images, factors such as \textbf{patch scale} and \textbf{anomaly type} can critically influence algorithm performance.
From a practical standpoint, the following strategies may be more applicable in clinical workflows:
\begin{enumerate}[label=\arabic*)]
    \item \textbf{Ensemble detection}: Apply multiple anomaly detection algorithms and flag regions with consistently high anomaly scores as suspicious.
    \item \textbf{Visualization}: Provide pixel-wise or patch-wise anomaly heatmaps to alert physicians, allowing them to make the final judgment.
    \item \textbf{Data-driven evolution}: As more abnormal samples are collected, expand the dataset and transition toward supervised approaches, which can achieve higher reliability and automation in computer-aided diagnosis.
\end{enumerate}
When selecting the training epoch, it may be helpful to monitor whether the training loss has converged. If computational resources permit, \textbf{sample-wise epoch selection} strategies can also be considered.}

\subsubsection{Outlook for Pathology-Specific Anomaly Detection}
To improve anomaly detection in pathology images, several future directions can be explored:
\begin{enumerate}[label=\arabic*)]
    \item \textbf{Domain-aware model design}: Incorporate domain-specific characteristics, such as \textbf{stain variability} and \textbf{scale diversity}, into model training. Building multiscale datasets and employing appropriate augmentation techniques are essential. Pretraining on data from the target imaging modality may also yield more relevant features.
    \item \textbf{Score aggregation and model fusion}: Aggregating anomaly scores across pixel-level, patch-level, and multiscale representations—and designing robust fusion strategies for multiple models—remain promising research directions.
    \item \textbf{Realistic evaluation and deployment considerations}
    This includes incorporating clinically relevant datasets beyond synthetic or toy examples, addressing practical challenges such as stain variation, limited abnormal annotations, and unbiased model selection. Developing algorithms with robustness to real-world variability is essential for reliable deployment in clinical settings.
\end{enumerate}

\subsection{Limitations}
This work has some limitations. To ensure consistent coverage between small-scale and large-scale patches, we used the default parameters provided in the program, which limited the number of large-scale patches used for training. Additionally, due to time and resource constraints, we were unable to fine-tune the model for each dataset. Instead, we used the default parameters uniformly across all datasets.

\section{Conclusions}
This review provides a extensive benchmark of 23 anomaly detection methods applied to digital pathology images, highlighting both their potential and limitations. Through evaluations on diverse real and synthetic datasets, we reveal that no single method consistently outperforms others, and performance is highly dependent on image scale, anomaly patterns, and training strategies. Our findings underscore the need for pathology-specific designs and robust evaluation protocols, offering valuable insights to guide future research in this emerging field.


\section*{Acknowledgment}
This work was supported by the National Institutes of Health under award numbers R01DK135597 (Huo), R01EB033385, R01DK132338, REB017230, R01MH125931, and R01DK128200 (KTW), DoD HT9425-23-1-0003 (HCY), and KPMP Glue Grant. This study was also supported by the National Science Foundation (2040462). This work was also supported by NSF NAIRR Pilot Award NAIRR240055. This project was also supported by The Leona M. and Harry B. Helmsley Charitable Trust grant G-1903-03793 and G-2103-05128. This research was also supported by Veterans Affairs Merit Review grants I01CX002662, I01CX002171 (KTW), and I01CX002473 (KTW). This work was also supported by Vanderbilt Seed Success Grant, Vanderbilt Discovery Grant, and VISE Seed Grant. We extend gratitude to NVIDIA for their support by means of the NVIDIA hardware grant.



%



\bibliography{reference} 
\bibliographystyle{IEEEtran} %

\end{document}